\documentclass[preprint,12pt]{elsarticle}

\usepackage{amssymb}
\usepackage{amsmath}
\usepackage{hyperref}
\hypersetup{
	colorlinks=true,
	linkcolor=blue,
	filecolor=magenta,      
	urlcolor=cyan,
}
\usepackage{lineno}
\journal{International Journal of Hydrogen Energy}
\begin{document}

\begin{frontmatter}
	
	\title{Confined combustion of lean hydrogen-air mixtures with suspended water micro-droplets}
	
	\author[label1]{I.S. Yakovenko \corref{cor1}}
	\ead{yakovenko.ivan@bk.ru}
	
	 \author[label1]{A.D. Kiverin}
	
	 \address[label1]{Joint Institute for High Temperatures, Izhorskaya st. 13 Bd.2, Moscow, 125412, Russia}
	\cortext[cor1]{Corresponding author. Joint Institute for High Temperatures, Izhorskaya st. 13 Bd.2, Moscow, 125412, Russia. Tel.: +7 4954844433}
	
	\date{\today}% It is always \today, today,
	
	\begin{abstract}
		The present study is devoted to the detailed numerical analysis of the combustion waves propagation through the lean hydrogen-air mixtures with suspended micro-droplets of water. Considered gaseous mixtures are characterized by a relatively low reactivity and so the intensity of the combustion waves is moderate. Herewith, it is found that the flame can be noticeably accelerated in the presence of suspended micro-droplets. In particular, it is shown that the interaction between the flame front and the droplets with a diameter larger than 25~$\mu$m results in the enhancement of the flame instability that provides a substantial acceleration of the mixture burn-out process. At increased number density of water droplets the combustion proceeds in a similar way to the combustion within a porous medium, and similar trends are observed such as the increase in flame speed and intensification of the combustion process with the pressure increasing inside the confined vessel. In such a way it can be concluded that the water should be atomized in a proper way when attempting to use it as hydrogen explosion inhibitor. Otherwise, water droplets could induce flame acceleration via the mechanisms distinguished in this paper.
	\end{abstract}

	
	\end{frontmatter}
	
%%	\modulolinenumbers[5]
	
%%	\begin{linenumbers}
	
	\section{\label{sec:level1} Introduction}
	Safe operation of nuclear power plants and perspective hydrogen energy systems is of primary importance for the development of the energy sector. One of the possible scenarios of the emergency on a nuclear power plant is the ignition of the combustible gaseous mixtures on the base of hydrogen that are formed under containment as a result of zirconium oxidation by water steam \cite{Udagawa2010}. Similar conditions could arise as a result of a hydrogen leak during its transfer or storage as a fuel \cite{Petukhov2009,Eichert1986}. In both cases, the released hydrogen mixes with ambient air forming a highly reactive combustible mixture. As a result of subsequent combustion, variety of different flame propagation modes can be realized. These modes are characterized by different dynamic and thermal loads from the slow diffusive combustion in the form of flame kernels in lean mixtures \cite{Ronney1990,Yakovenko2017} to the devastating trans- or supersonic flames or even detonation in near stoichiometric mixtures \cite{Ciccarelli2008,Ivanov2011,Kiverin2016}. Thereby the problems of ignition and flame development in hydrogen-based gaseous mixtures require the detailed analysis for the elaboration of robust approaches for effective mitigation of the flame acceleration or its complete extinguishing. One of the possible ways for the complete or partial mitigation of the combustion is the addition of the dispersed droplets of water into the formed combustible mixture. Due to the low cost, high specific heat capacity and latent heat, water is one of the widely used combustion inhibitors \cite{Grant2000}. 
	
	Suppression of fire propagation by water droplets is determined by several physical mechanisms such as flame cooling, cooling of the fuel, dilution of the fuel with inert water vapor and others. Thus, the effectiveness of those mechanisms depends differently on the average droplet flow related to the unit of the flame front surface, droplets size and spatial distribution, droplets velocity relative to the flame front \cite{ZhigangLiu1999}. Set of parameters for successful flame mitigation can be determined individually and depends on gas-dynamical conditions, fuel characteristics and flame front structure. In the optimal case, the residence time of the droplets inside the flame front should be maximized \cite{Thomas2000}. Droplets of the small size have a large ratio of the surface area to the volume, so small droplets evaporate faster and release more water vapor per time unit than large droplets do. Thereby small droplets provide more effective cooling and dilution of the combustible mixture inside the reaction zone. On the other hand, small droplets are less inert and the flow ahead of the flame can prevent access of droplets to the flame front. Larger droplets are more inert, evaporate slower and can be effectively used for the cooling of the combustion products. However, that does not always lead to flame quenching. Altogether experiment shows that the quenching of a deflagrative or even detonative combustion \cite{Thomas1991,Thomas1990}, the attenuation of a blast wave \cite{VanWingerden1995}, the decrease in flame acceleration rate and the delay in the deflagration-to-detonation transition process \cite{Boeck2015} are possible at the optimal set of parameters of the dispersed water phase. In \cite{Tsai1982} the increase in the low flammability limit of the hydrogen-air mixture from 4\% to 4.5--5.3\% of H$_2$ for droplets of 20--50~$\mu$m diameter and to 7.2--8.5\% for droplets of 10~$\mu$m was observed. In \cite{Medvedev2002} the presence of droplets of 1--2~$\mu$m diameter formed in the process of water vapor condensation in hydrogen-air mixture resulted in low flammability limit increase up to 15\% of H$_2$.
	
	Despite the recognized ability of the water droplets to suppress the combustion process, water droplets admixture to the combustible compound does not always lead to the desired outcome. Thus, it is known that the injection of the water spray can cause a large scale flow turbulization in the combustible mixture \cite{VanWingerden1995_2,Thomas1982}. The interaction between the flow generated in the fresh mixture by the expanding combustion products and the droplets is also a possible way of flow turbulization \cite{Thomas1982}. However, the latter mechanism seems to be relevant only for the fast flames subjected to the large droplets \cite{VanWingerden1995_2}. Flow turbulization ahead of the flame can be a reason for the intensification of the flame surface wrinkling that results in the opposite effect - flame accelerates instead of being suppressed \cite{Gieras2008,Zhang2014,Cheikhravat2015}. Another possible mechanism of the flame acceleration in the presence of water droplets is the enhancement of the flame surface instability development. Thus, recent numerical analysis has shown that suspended droplets in the premixed combustible gaseous mixture can trigger Darrieus-Landau instability and provoke the generation of small-scale perturbations on the flame surface \cite{Nicoli2017,Nicoli2019}. As a result, the flame front surface becomes corrugated that can significantly increase the flame propagation velocity. 
	
	Considering the possible application of fine water droplets for fire extinguishing in closed compartments \cite{Joseph-Auguste2009}, it is important to determine the mechanisms of the droplets influence on the flame propagation dynamics and the flame front spatial structure in the confined vessels. The present paper is devoted to the numerical analysis of the effect of fine water droplets presence on the confined flame propagation in the hydrogen-based gaseous mixture.     
	
	\section{\label{sec:level2} Problem setup and numerical method}
	
\begin{figure}
		\centering\includegraphics[width=0.65\linewidth]{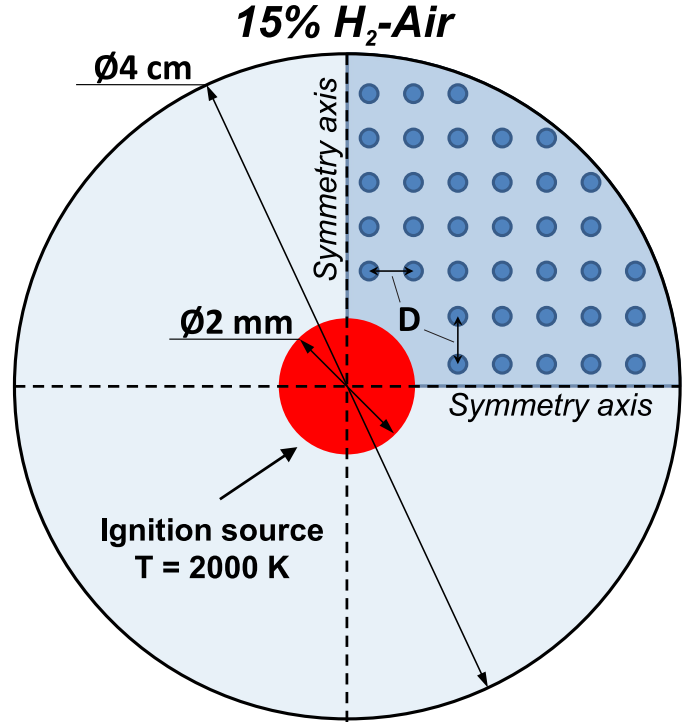}
		\caption{\label{figure1} Computational domain setup.}
	\end{figure}

	Flame propagation was studied in the confined two-dimensional vessel with diameter 4~cm filled with the lean 15\% H$_2$-air mixture under normal conditions ($p_0$ = 1~atm, $T_0$ = 300~K). Herewith, the gaseous mixture was seeded with water micro-droplets with the diameter $d_d$ in the range of 25-200~$\mu$m. Problem setup is schematically presented in Fig.~\ref{figure1}. Initially, water droplets were located in the array with the distance between neighboring droplets $D$ that was varied in the range between 0.4~mm and 3.0~mm. The mixture was ignited in the gas volume heated up to 1500~K in the middle of the vessel. To avoid the influence of water droplets on the ignition stage, droplets were not placed in the middle area of the vessel. Symmetry conditions were imposed on the left and bottom boundaries. Adiabatic wall boundary was imposed on the outer wall of the vessel. 
	
	Full Navier-Stokes system of the conservation laws was considered in the low-Mach number approximation to reproduce the combustion dynamics. Thermal conductivity, multicomponent diffusion and energy release associated with the chemical transformations were taken into account. The solution was obtained for the following system of the governing equations \cite{Kuo}:
	
	\begin{equation}
	\frac{{\partial \rho }}{{\partial t}} + \frac{{\partial \rho {u_i}}}{{\partial {x_i}}} = \dot m  
	\label{eq:1}
	\end{equation}
	
	\begin{equation}
	\frac{{\partial \rho {Y_k}}}{{\partial t}} + \frac{{\partial \rho {Y_k}{u_i}}}{{\partial {x_i}}} = \nabla \rho {Y_k}{V_{k,i}} + {\dot \omega _k} + {\dot m_k}
	\label{eq:2}
	\end{equation}
	
	\begin{equation}
	\frac{{\partial \vec u}}{{\partial t}} - \vec u \times \vec \omega  + \nabla H - \tilde p\nabla \left( {\frac{1}{\rho }} \right) = \frac{1}{\rho }\left[ {{{\vec f}_b} + \nabla \sigma } \right]
	\label{eq:3}
	\end{equation}
	
	\begin{equation}
	\frac{{\partial \rho {h_s}}}{{\partial t}} + \frac{{\partial \rho {h_s}{u_i}}}{{\partial {x_i}}} = \frac{{d\bar p}}{{dt}} - \sum\limits_{k = 1}^N {{{\dot \omega }_k}\Delta h_{f,k}^0}  - \frac{\partial }{{\partial {x_i}}}\left( {\rho \sum\limits_{k = 1}^N {{h_{s,k}}{Y_k}{V_{k,i}}} } \right) - \frac{\partial }{{\partial {x_i}}}\left( {\kappa \frac{{\partial T}}{{\partial {x_i}}}} \right) + {\sigma _{ij}}\frac{{\partial {u_i}}}{{\partial {x_j}}} + \dot q
	\label{eq:4}
	\end{equation}
	
	\begin{equation}
	{\sigma _{ij}} = \mu \left( Y_k, T \right)\left[ {\frac{{\partial {u_i}}}{{\partial {x_j}}} + \frac{{\partial {u_j}}}{{\partial {x_i}}} - \frac{2}{3}{\delta _{ij}}\frac{{\partial {u_l}}}{{\partial {x_l}}}} \right]
	\label{eq:5}
	\end{equation}
	
	\begin{equation}
	\bar p = \rho RT\sum\limits_k {\frac{{{Y_k}}}{{{M_k}}}}
	\label{eq:6}
	\end{equation}
	
	\begin{equation}
	dh_s = C_p\left(Y_k,T\right) dT
	\label{eq:7}
	\end{equation}
	
	Here $\rho$ is the mass density, $\vec u$ is the mass velocity, $u_i$ are the mass velocity vector components, $Y_k$ is the molar fraction of $k$-th component of gaseous mixture, $\tilde p$ is the dynamic component of pressure fluctuations, which is by the order of magnitude much smaller compare with thermodynamic pressure $\bar p$ and $p\left( {\vec x,t} \right) = \bar p\left( t \right) + \tilde p\left( {\vec x,t} \right)$, $\sigma$ is the viscous stress tensor, $\sigma _{ij}$ are the components of the viscous stresses tensor, $H = \left| {{{\vec u}^2}} \right|/2 + \tilde p/\rho $ is the stagnation energy per unit mass, $\vec \omega$ is the vorticity vector, $h_s$ is the specific sensible enthalpy of the mixture, $h_{s,k}$ is the specific sensible enthalpy of $k$-th component of gaseous mixture, $T$ is the temperature, $\kappa \left(Y_k, T\right)$ is the thermal conductivity coefficient, $\mu \left(Y_k, T\right)$ is the viscosity coefficient, $h_{f,k}^0$ is the enthalpy of formation of $k$-th component of the gaseous mixture, $\vec V_{k,i}$ is the diffusion velocity vector component of $k$-th specie, $C_p(Y_k,T)$ is the specific heat capacity at constant pressure of the mixture. Term $\dot \omega _k$ represents the change in molar fraction of $k$-th specie due to the chemical reactions. $\dot m$, $\dot m_k$, $\dot q$ are respectively the changes in mass, mass fraction of $k$-th species and energy change associated with movement and evaporation of droplets, ${\vec f}_b$ is the droplets drag force. 
	
	To resolve droplets dynamics Lagrangian model was implemented for individual droplets \cite{McGrattan}. Droplets velocities ${\vec u}_d$ and coordinates ${\vec x}_d$ were found from the following equations of motion: 
	
	\begin{equation}
	\frac{{d{{\vec u}_d}}}{{dt}} =  - \frac{1}{2}\frac{{\rho {C_d}{A_{d,c}}}}{{{m_d}}}\left( {{{\vec u}_d} - \vec u} \right)\left| {{{\vec u}_d} - \vec u} \right|
	\label{eq:8}
	\end{equation}
	
	\begin{equation}
	\frac{{d{{\vec x}_d}}}{{dt}} = {\vec u_d}
	\label{eq:9}
	\end{equation}
	
	Droplets drag force was given by the relation:
	
	\begin{equation*}
		{\vec f_b} = \frac{1}{V}\sum\limits_{droplets} {\left[ {\frac{{{\rho }}}{2}{C_d}{A_{d,c}}\left( {{{\vec u}_d} - \vec u} \right)\left| {{{\vec u}_d} - \vec u} \right| - \frac{{d{m_d}}}{{dt}}\left( {{{\vec u}_d} - \vec u} \right)} \right]} 
	\end{equation*}
	
	Drag coefficient $C_d$ was modeled via empirical relation taking into account character of the flow:
	
	\begin{equation*}
		{C_d} = \left\{ \begin{array}{l}
			24/{{{\rm Re}} _D}\;,\;{{{\rm Re}} _D} < 1\\
			24\left( {0.85 + 0.15{{\rm Re}} _D^{0.687}} \right)/{{{\rm Re}} _D},\;1 < {{{\rm Re}} _D} < 1000
		\end{array} \right.
	\end{equation*}
	here ${{{\rm Re}} _D} = \frac{{\rho \left( {{{\vec u}_d} - \vec u} \right){d_d}}}{{\mu \left(Y_k,T \right)}}$ is the droplet-based local Reynolds number. 
	
	Droplets heating and evaporation and corresponding inter-phase exchange was modeled via the following equations:
	
	\begin{equation}
	\frac{{d{m_d}}}{{dt}} =  - {A_{d,s}}{h_m}{\rho}\left( {{Y_{k,l}} - {Y_k}} \right)
	\label{eq:10}
	\end{equation}
	
	\begin{equation}
	{\rho _g}V\frac{{d{Y_k}}}{{dt}} =  - \frac{{d{m_d}}}{{dt}}
	\label{eq:11}
	\end{equation}
	
	\begin{equation}
	\frac{{d{T_d}}}{{dt}} = \frac{1}{{{m_d}{C_{p,d}}}}\left[ {{A_{d,s}}h\left( {T - {T_d}} \right) + \frac{{d{m_d}}}{{dt}}{h_v}} \right]
	\label{eq:12}
	\end{equation}
	
	\begin{equation}
	\frac{{dT}}{{dt}} = \frac{1}{{{m_g}{C_{p}}}}\left[ {{A_{d,s}}h\left( {{T_d} - T} \right) + \frac{{d{m_d}}}{{dt}}\left( {{h_v} + {h_l}} \right)} \right]
	\label{eq:13}
	\end{equation}
	
	Hence, 
	
	\begin{equation*}
		\dot m = \frac{1}{V}\frac{{d{m_d}}}{{dt}}
	\end{equation*}
	\begin{equation*}
		\dot q = \frac{1}{V}\left[ {{A_{d,s}}h\left( {{T_d} - T} \right) + \frac{{d{m_d}}}{{dt}}\left( {{h_v} + {h_l}} \right)} \right]
	\end{equation*}
	here $Y_{k,l}$ is the liquid equilibrium mass fraction of vapor, $T_d$ is the droplet temperature, $m_d$ is the mass of the individual droplet, $A_{d,s}$ is the surface area of the droplet, $h_m$ is the mass transfer coefficient, $C_{p,d}$ is the liquid specific heat, $h$ is the heat transfer coefficient between the droplet and the gas, $h_l$ is the liquid specific enthalpy, $h_v$ is the latent heat of vaporization of the liquid. Mass transfer and heat transfer coefficients are described with empirical correlations:
	\begin{equation*}
		h_m = \frac{\mathrm{Sh} \cdot D_{lg}}{L}
	\end{equation*}
	\begin{equation*}
		h = \frac{\mathrm{Nu} \cdot \kappa\left(Y_k,T\right)}{L}
	\end{equation*}
	here $D_{lg}$ is the diffusion coefficient of liquid (water) vapor and gaseous mixture. Sherwood and Nusselt numbers were obtained from the empirical laws for the spherical droplets:
	${{\rm Sh} = 2 + 0.6{\rm Re} _D^{1/2}{\rm Sc}^{1/3}}$, 
	${{\rm Nu} = 2 + 0.6{\rm Re} _D^{1/2}{\rm Pr}^{1/3}}$. 
	
	Specific heat capacities and enthalpies of formation were calculated according to the JANAF tables \cite{Chase1998}. Diffusion model is based on the zeroth-order Hirshfelder-Curtiss approximation \cite{HirschfelderCurtissBird1964}. Mixture averaged transport coefficients were obtained from the first principles of the gas kinetics theory \cite{KeeColtrinGlarborg2003}. Diffusion velocities were calculated taking into account correction velocity approach proposed in \cite{Coffee1981}. 
	
	The governing equations \ref{eq:1}-\ref{eq:7} were solved using the second-order predictor/corrector method implemented in \cite{McGrattan}. The equations of motion of the Lagrangian droplets \ref{eq:8}-\ref{eq:9} along with the equations describing heat and mass transfer between the liquid and gas \ref{eq:10}-\ref{eq:13} were solved via semi-explicit method proposed in \cite{McGrattan} as well.
	
	Chemical kinetics was modeled using the contemporary kinetic scheme presented in \cite{Keromnes2013}. The system of ODEs defining the combustion kinetics was solved by Gear method. 
	
	\section{\label{sec:level3} Results and discussion}
	
	Outwardly propagating flame is subjected to the development of the hydrodynamic Darrieus-Landau instability and thermo-diffusive instability \cite{Matalon2007}. In the considered case of the lean hydrogen-air mixture with 15\% H$_2$ content, the instability development begins immediately after the ignition as the Lewis number is sufficiently less than unity, $\mathrm{Le} \approx 0.6$ (according to our estimations and the calculations from \cite{Law2006}). Due to the negative Markstein number \cite{safekinex}, the rate of the flame surface growth due to instability is rather high, so one can observe a developed cellular flame structure already on the initial stages of the process \cite{Bauwens2017}. The expansion of the combustion products leads to the formation of the flow ahead of the flame in the fresh mixture. If the flame is confined, then the flow pushed by the propagating flame reflects from the confinement boundaries and interacts with the flame. That can trigger the thermo-acoustic instability of the flame \cite{Akkerman2013,Al-Shahrany2005} that supplements the Darrieus-Landau and thermo-diffusive instabilities development. Thereby the overall process of the confined flame evolution in the lean hydrogen-air mixture is rather complex.
	
	\begin{figure}
		\centering\includegraphics[width=1.0\linewidth]{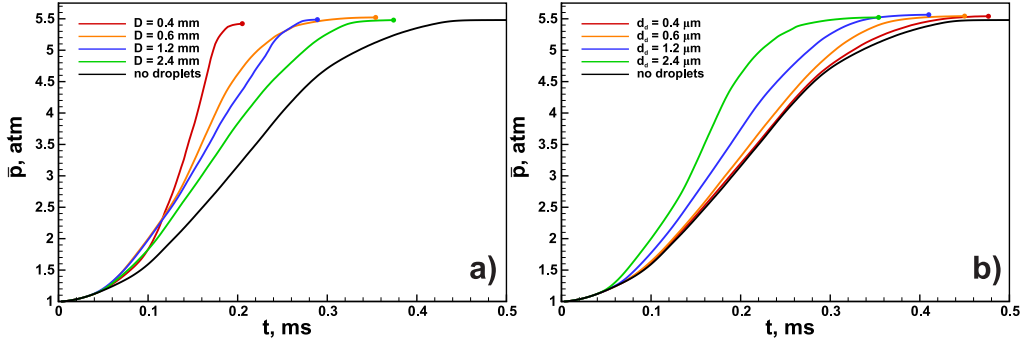}
		\caption{\label{figure_2} Time histories of the pressure rise in the mixtures with suspended micro-droplets a) with the diameter $d_d = 200$~$\mu$m and initial distance between droplets $D$ = 0.4--2.4~mm; b) with the diameter $d_d = 25$, 50, 100, 200~$\mu$m and initial distance between droplets $D = 0.6$~mm. Pressure curve of the pure gaseous mixture is given for the comparison. Dots represent the time instant when the combustion is complete.}
	\end{figure}
	
\begin{figure}
		\centering\includegraphics[width=0.75\linewidth]{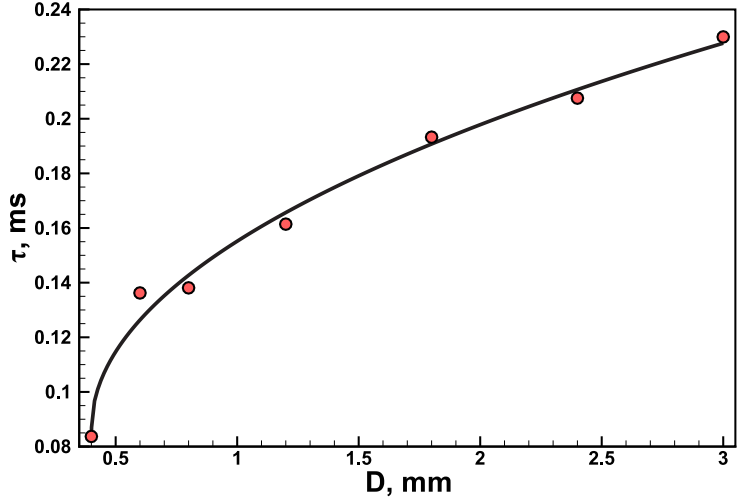}
		\caption{\label{figure_3} Characteristic burn-out time $\tau$ vs initial distance between droplets $D$. Droplets diameter $d_d=200$~$\mu$m. Solid line represents the best algebraic curve fitting.}
	\end{figure}

	One of the main characteristics of the combustion in the confined vessel is the rate of the pressure rise in the volume. Thus, the rate of the pressure rise indicates the intensity of the combustion process and can be used to estimate the laminar burning velocity in closed vessel explosions experiments \cite{Dahoe2005}. In Figure~\ref{figure_2}a the pressure-time histories of the combustion process are given for the different configurations of the initial array of droplets of 200~$\mu$m diameter, namely for the different droplet inter-distance $D$. One can notice that the combustion intensifies in the presence of water droplets. Moreover, the effect is more pronounced for the higher droplets number density (shorter droplets inter-distance). Herewith, in the case of larger droplets combustion proceeds faster than in the case of smaller droplets (Figure~\ref{figure_2}b). To describe the combustion intensity inside the confined volume from the integral point of view one can introduce the following integral time-scale parameter $\tau=\frac{\left(p_f-p_0\right)}{\left(dp/dt\right)_{max}}$, where $p_0$ is the initial pressure and $p_f$ is the pressure when the combustion process is complete. Dependence of the characteristic burn-out time scale $\tau$ on the initial distance between droplets of 200~$\mu$m diameter is presented in Figure~\ref{figure_3}. 
	%\begin{figure}
	%\includegraphics[width=30pc]{fig4bx.eps}
	%\caption{\label{figure_2b} Pressure rise histories of the mixture with suspended micro-droplets with the diameter $d_d$ = 25, 50, 100, 200~$\mu$m and initial distance between droplets $D$ = 0.6~mm. Pressure curve of the pure gaseous mixture is given for the comparison. Dots represent the time instant when the combustion is complete.}
	%\end{figure}
	
\begin{figure}
		\centering\includegraphics[width=1.0\linewidth]{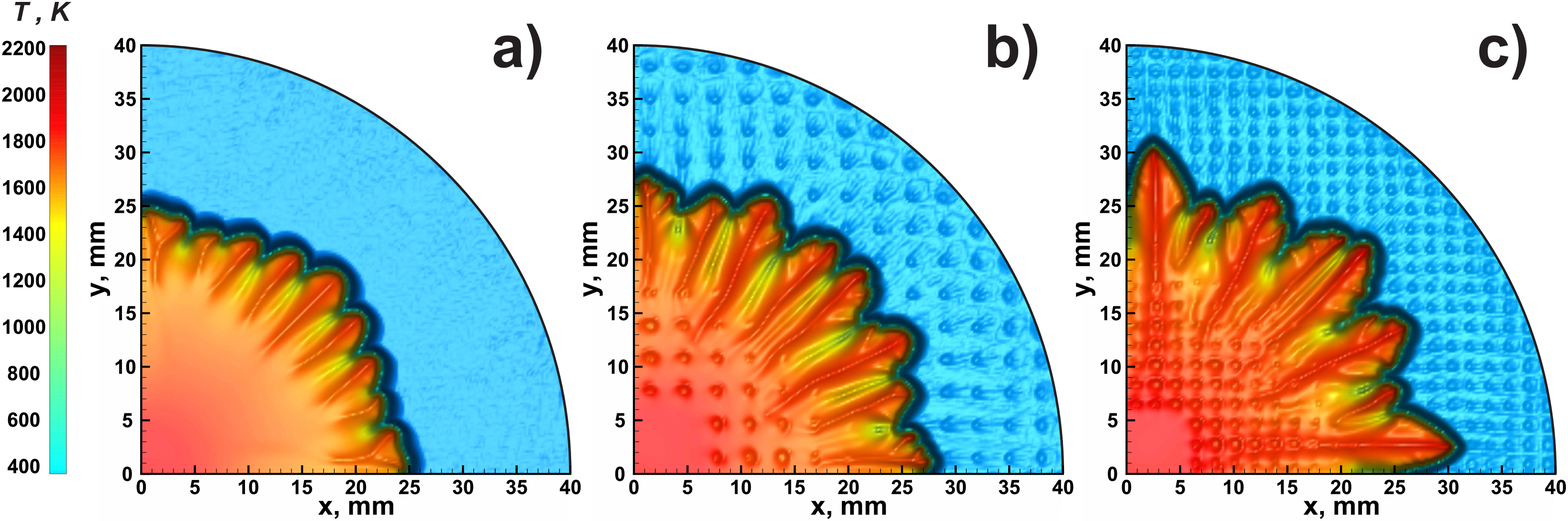}
		\caption{\label{figure_4} Spatial structure of the flame in the pure 15\%H$_2$-air gaseous mixture (a) and in the gaseous mixture with suspended micro-droplets of water with diameter 200~$\mu$m and the initial distance between droplets 3~mm (b) and 1.8~mm (c). Time instant 10~ms.}
	\end{figure}

	Flame surface growth due to the instabilities development leads to the flame acceleration above the laminar flame speed values. Addition of the micro-droplets of liquid can also lead to instability enhancement and the increase in the flame propagation velocity \cite{Nicoli2017,Nicoli2019}. Liquid droplets act like obstacles that cause momentum and energy losses and force the flame to bend while flowing around the droplet in a similar way as in presence of solid obstacles \cite{Ogawa2013}. Perturbations caused by the interaction between the flame and droplets give rise to the instabilities and the formation of highly corrugated flame. In Figure~\ref{figure_4}, the structures of the flame in the pure gaseous mixture and the mixture with suspended micro-droplets of water ($d_d = 200$~$\mu$m) are presented. From the comparison of the flame front structures, one can conclude that the droplets effect on the flame surface increases when the distance between droplets is shorter. Thus, the combustion process intensifies as droplets inter-distance decreases (Fig.~\ref{figure_2}a). 
	
	The instability of the flame front is characterized by the wave-lengths that grow faster than others. According to the linear stability analysis, in the case of Darrieus-Landau instability the shortest, critical wave-lengths of the order of the flame front thickness $L_f$ have the fastest growth rate \cite{Matalon2007}. Critical wave-length can be estimated via relation proposed by \cite{Zaytsev2002} as ${\lambda_c = \frac {2 \pi \chi}{u_f} \left( 1 + \theta \frac{\theta + 1}{\left(\theta-1\right)^2}\mathrm{ln}\theta\right)}$, where $\theta$ is the expansion factor that is the ratio of density of the fresh mixture $\rho_a$ and density of the combustion products $\rho_b$, $\chi$ is the thermal diffusivity of the mixture, $u_f$ is the laminar flame speed. The perturbations imposed by the droplets can be characterized by the spatial scale $D$. Thus the droplets have the greatest impact on the flame development if they excite the fastest-growing perturbations on the flame front, namely when $D \sim \lambda_c$. 
	
\begin{figure}
		\centering\includegraphics[width=0.75\linewidth]{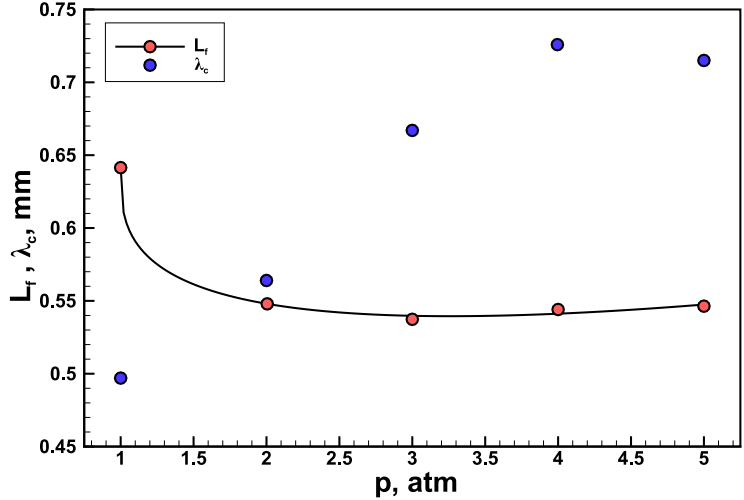}
		\caption{\label{figure_5} Flame front thickness $L_f$ (red dots) and critical wave-length $\lambda_c$ (blue dots) dependence on the pressure of the 15\%H$_2$-air gaseous mixture. Line represents best algebraic curves fitting.}
	\end{figure}

\begin{figure}
		\centering\includegraphics[width=0.75\linewidth]{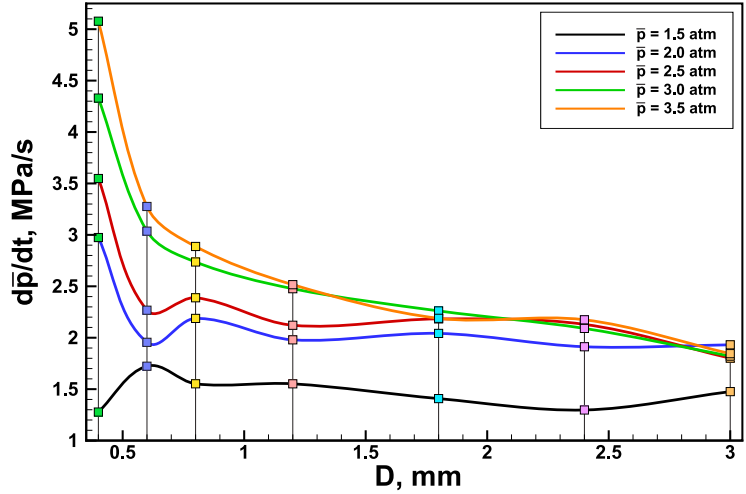}
		\caption{\label{figure_6} Dependencies of the pressure rise $d\bar p/dt$ on the initial distance between droplets $D = 0.4$--2.4~mm for different pressure values in the process of the mixture burn-out (Fig.~\ref{figure_2}a). Droplets diameter -- 200~$\mu$m.}
	\end{figure}

	A set of 1D planar flame calculations was carried out beforehand to assess the flame front thickness $L_f =\frac{\left(T_b-T_a\right)}{|\left(dT/dx\right)_{max}|}$, where $T_b$ is the temperature of the combustion products, $T_a$ is the temperature of the fresh mixture, and critical wave-length $\lambda_c$ at different pressure values. According to the results, flame front thickness $L_f$ of the lean 15\% H$_2$-air mixture under normal conditions is equal to 0.64~mm (Fig.~\ref{figure_5}) and decreases with pressure, while critical wave-length $\lambda_c$ equals to 0.49~mm and grows with pressure. Thus, the setup with droplet array configuration with $D \approx 0.49$~mm should provide the highest combustion intensity on the initial stage of the process. To determine the instantaneous combustion intensity one can analyze the behavior of the pressure derivative $d\bar p/dt$ for different pressure values achieved at consequent stages of combustion. The dependence of $d\bar p/dt$ on the initial distance between droplets $D$ is presented in Figure~\ref{figure_6} for pressures 1.5, 2.0, 2.5, 3.0 and 3.5~atm. One can notice that for the pressure 1.5~atm the value of derivative $d\bar p/dt$ has a maximum for $D=0.6$~mm that is in line with our reasoning. However, as the pressure rises, the critical wave-length $\lambda_c$ becomes larger (Figure~\ref{figure_5}) and perturbations imposed by the droplets configuration with $D=0.8$~mm start to intensify the process more effectively. Due to this, a local maximum of $d\bar p/dt$ can be observed at $D\sim 0.8$~mm at rising pressure ($p=2.0$ and 2.5~atm, Fig. \ref{figure_5}).
	
	It should be noted that in the framework of the considered model of interaction between droplets and gaseous phase, droplets impose energy and momentum losses, while the effects of flow around the surface of droplet and droplets coalescence are neglected. In the case of the smallest distance between droplets $D=0.4$~mm the suspension consisting of the fresh mixture and water droplets can be considered as a porous medium and the combustion propagation mechanism switches to one similar to the filtration combustion \cite{Babkin1993}. It is well known that in such a system different modes of combustion propagation can be observed, including high-velocity combustion with the velocity up to several times higher than the laminar flame speed \cite{Babkin2011,Pinaev1995}. The presence of losses, in particular, thermal losses and hydraulic resistance, is considered as the reason for combustion amplification in such kind of systems \cite{Brailovsky2000,Kagan2010}. In the case of suspension with $D=0.4$~mm, which is smaller than the flame front thickness, droplets act as a porous matrix providing sufficient hydraulic resistance for the flame acceleration. Another characteristic feature of the filtration combustion that was observed experimentally is the increase of the flame speed with pressure \cite{Babkin1991}. In the case of the confined combustion, a gradual increase of pressure inside the chamber due to mixture burnout can cause further enhancement of filtration combustion. Apparently, those effects are responsible for the obtained increase of the pressure rise intensity in the system with the smallest droplet inter-distance $D=0.4$~mm, although a more precise model of inter-phase interaction should be implemented for a deeper understanding.
	
	\begin{figure}
		\centering\includegraphics[width=1.0\linewidth]{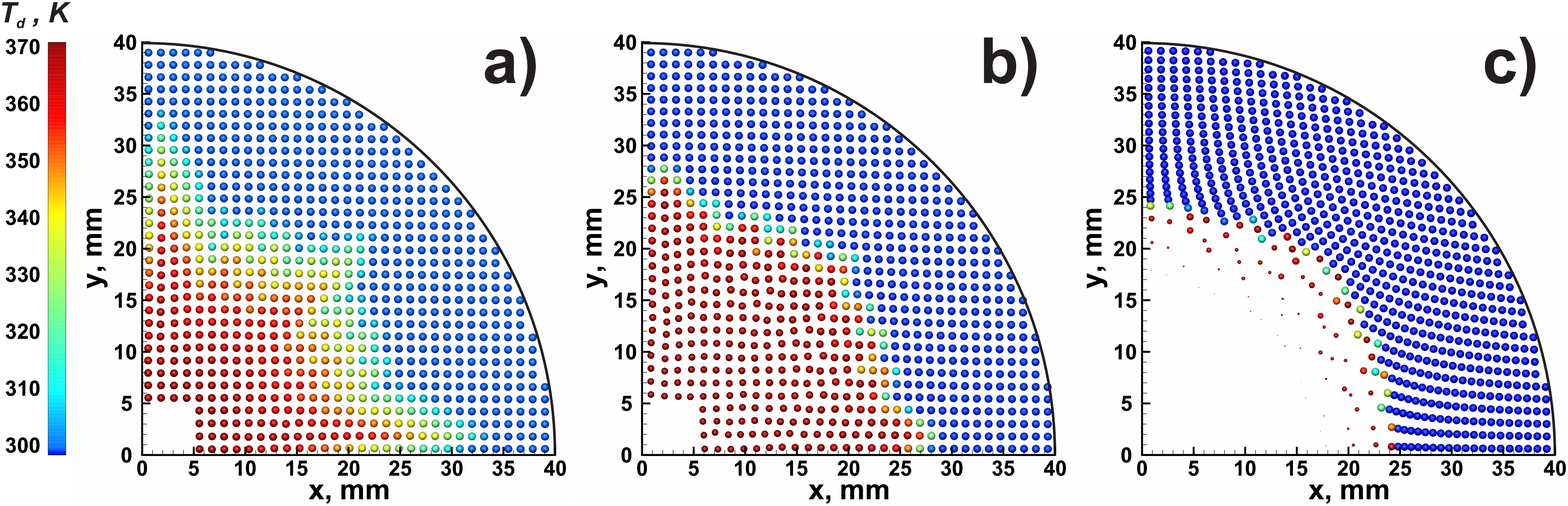}
		\caption{\label{figure_7} Snapshots of the droplets location. Initial distance between droplets $D = 1.2$~mm. Droplets diameter (a) 200~$\mu$m, (b) 100~$\mu$m, (c) 25~$\mu$m. Markers represent droplet locations, marker sizes are normalized by the initial mass of the droplet. Time instant 10~ms.}
	\end{figure}
	
	So far we have mainly considered the droplets with diameter $d_d=200$~$\mu$m that are characterized by the greatest influence on the confined combustion process among the examined cases (Fig.~\ref{figure_2}b). The behavior of the individual droplets of different diameters $d_d$ can be seen from the Figure~\ref{figure_7}. The rate of heating of the droplet depends linearly on the surface of the droplet that is proportional to the square of the droplet diameter $~d_d^2$. Thus, larger droplets are heating and evaporating slower than the smaller ones. As it can be seen from the Figure~\ref{figure_7} the droplets heating front is much wider for the $d_d=200$~$\mu$m, while in the case of $d_d=100$~$\mu$m and $d_d=25$~$\mu$m the droplets become heated to the critical temperature immediately inside the flame front and the smallest droplets $d_d=25$~$\mu$m are evaporating intensively in the region of the combustion products. Heat losses due to the droplets evaporation and combustible mixture dilution by the water vapor are the main mechanism of the combustion suppression and quenching by the water droplets. However, no quenching was observed even for the smallest droplets of diameter $d_d=25$~$\mu$m. In the all considered cases, evaporation took place mainly in the region of the combustion products and did not affect the reaction zone explicitly. Though the value of the final pressure $p_f$ is slightly increased due to the release of water during the flame propagation. 
	
	One can notice that for the largest droplets among the considered ones the structure of the initial droplets array is almost unchanged throughout the process (Figure~\ref{figure_7}a). As the flow in the fresh mixture reflects from the vessel walls, the resultant flow velocity ahead of the flame is rather small. Thus, the influence of the flow in the fresh mixture on the droplets is greatly reduced, so the larger droplets with sufficient inertia remain almost static. Although smaller droplets are carried by the flow, they still can interact with the flame surface. This indicates that the ability of water droplet to suppress the flame can vary depending on the features of the flows generated during the combustion process in the particular geometry of the vessel.

	\section{\label{sec:level4} Conclusions}
	In the present paper the confined combustion of the hydrogen-air mixture in the presence of volumetrically suspended water micro-droplets is considered employing numerical analysis. It is shown that the admixture of water micro-droplets can lead to the substantial acceleration of the burn-out process. Analysis of the combustion dynamics allowed to conclude that one of the main roles in combustion enhancement belongs to the development of the flame front instability. Thus, the combustion proceeds with greater amplification when the droplets inter-distance becomes close to the critical wave-length $\lambda_c$ of Darrieus-Landau instability, so the droplets excite fastest-growing perturbations on the flame surface. The observed effect is most pronounced for larger droplets (of 200~$\mu$m diameter). Due to the small inertia of smaller droplets of 100$~\mu$m and 25$~\mu$m diameter, the induced momentum losses are lower and the generation of flame surface perturbations is less effective. Moreover, small droplets tend to be carried away by the flow formed ahead of the flame, so their influence on the flame structure and dynamics is reduced. In the case of high number density of droplets (smallest distance between droplets), the combustion proceeds in a similar way to the combustion within a porous media, and similar trends are observed such as the increase in flame propagation velocity and amplification of the combustion process with the pressure increase. Thus, the basic mechanisms responsible for the combustion intensification in presence of water micro-droplets are distinguished that can be used to improve the practical approaches of hydrogen explosion mitigation by water spraying. In particular, in view of obtained results it can be recommended to atomize the water in a proper way when attempting to use it as hydrogen explosion inhibitor. Otherwise, water droplets could induce flame acceleration via the mechanisms distinguished in this paper.
	
	\section*{\label{sec:level5} Acknowledgements}
	The research was financially supported by the Russian Foundation for Basic Research (grant №18-38-20079). Computational modeling was performed via computational software developed with the support of the Program of the Presidium of RAS ``Mechanisms of fault tolerance assurance of the contemporary high-performance and high-reliable computations''. We acknowledge high-performance computing support from the Joint Supercomputer Center of the Russian Academy of Sciences and Supercomputing Center of Lomonosov Moscow State University.
	
%%	\end{linenumbers}

	\section*{References}
	\bibliography{bibliography}% Produces the bibliography via BibTeX.

\end{document}